\documentclass[twocolumn,aps,prd,amsmath,amssymb,floatfix,showpacs]{revtex4}

\usepackage{bm}
\usepackage{amsmath}
\usepackage{epsf}

\newcommand{\higgs}{$\left<\phi\right>$\rm\ }

\newcommand{\ApJL}{Astrophys. J. Lett.}
\newcommand{\ApJ}{Astrophys. J.}

\newcommand{\PRD}{Phys. Rev. D}
\newcommand{\PR}{Phys. Rep.}

\newcommand{\AsAs}{Astron. Astrophys.}
\newcommand{\NAT}{Nature (London)}

\newcommand{\aut}[2]{{#2\ #1,}}
\newcommand{\refs}[6]{#2, {#3},  {#4} (#5).}

\begin{document}

\title{Big Bang nucleosynthesis and cosmic microwave background
constraints on the time variation of the Higgs vacuum expectation value}

\author{Jerry Jaiyul Yoo}
\thanks{jaiyul@astronomy.ohio-state.edu}
\address{Department of Astronomy, The Ohio State University, 
Columbus, OH 43210}

\author{Robert J. Scherrer}
\thanks{scherrer@pacific.mps.ohio-state.edu}
\address{Department of Physics, The Ohio State University,
Columbus, OH 43210\\
and Department of Astronomy, The Ohio State University, 
Columbus, OH 43210}

\begin{abstract}
\baselineskip 11pt
We derive constraints on the time variation of the Higgs vacuum expectation value
$\left<\phi\right>$ through the effects on
Big Bang nucleosynthesis (BBN) and the cosmic microwave background (CMB).
In the former case, we include the (previously-neglected) effect of
the change in the deuteron binding energy, which alters both the
$^4$He and deuterium abundances significantly.
We find that the current BBN limits on the relative change in
\higgs are 
$-(0.6 - 0.7) \times 10^{-2} < \Delta\!\!\left<\phi\right>\!/\!\left<\phi\right>
<
(1.5 - 2.0) \times 10^{-2}$, where the exact limits depend on
the model we choose for the dependence of the deuteron binding
energy on \higgs.
The limits from the current CMB data are much weaker.
\end{abstract}
\maketitle

\section{Introduction}
Physicists have long speculated that the fundamental constants
might not, in fact, be constant, but might instead vary with time \cite{dirac}.
Among the various possibilities the interaction coupling constants have
received the greatest attention (for a recent review,
see Ref. \cite{uzan}).
A great deal of attention has been focused on the fine
structure constant $\alpha$, for which a time variation
has been claimed in observations of quasar absorption lines \cite{fine}.
Cosmological limits on a time variation in $\alpha$ can be derived from
both big-bang nucleosynthesis
(BBN) and the cosmic microwave background (CMB) \cite{hannestad,manoj,berg,
ichikawa,avelino,martins,nollett}.  Comparatively less interest has been inspired
by possible time variation in the strength of the weak interaction,
as parametrized by the Fermi coupling $G_F$.

However, as emphasized by Dixit and Sher \cite{dixit_sher}, the Fermi constant
is not a fundamental constant in the
same sense as the fine structure constant.
The Fermi constant is given by
\begin{equation}
\label{G_F}
{{G_F}\over{\sqrt2}}={1\over{2\left<\phi\right>^2}}.
\end{equation}
Therefore, it is more appropriate to examine the variation of the Higgs
vacuum expectation value than that of the Fermi constant.
In fact, the time variation of the vacuum expectation value of a field
seems somewhat more plausible than the time variation of a fundamental
coupling constant.  (For arguments in favor of a possible {\it spatial}
variation of the Higgs vacuum expectation value, see Ref. \cite{agrawal}).

Limits on the time variation of the Higgs vacuum expectation value
have been derived from BBN \cite{dixit_sher,ss,ichikawa}.
However, improved observational limits on the primordial
element abundances now allow us to place
stronger limits than those derived, for example in Ref. \cite{ss}.
Furthermore, these previous studies ignored the effect of
the change in the deuteron binding energy, which we have incorporated
into our calculations.

The effects of a variation in \higgs on the CMB have been investigated
previously in Ref. \cite{kujat}, but this paper made no
attempt to derive actual limits based on observational CMB data, which
we will do here.

In this paper, we derive constraints on the time variation of \higgs
from both BBN and CMB observations.
In this regard, the present work is most similar in spirit to that of
Avelino et al. \cite{avelino}, who did a comparable calculation for
$\alpha$.  As noted in Ref. \cite{kujat}, the effect on the CMB
of changing the Higgs vacuum expectation value is similar, but not quite
identical, to the effect of changing $\alpha$.  The effects on BBN, on
the other hand, are quite different.  We assume that \higgs has
the same value at the era of primordial nucleosynthesis
$T \sim 10^{10} - 10^9$ K as it has at recombination $T \sim 10^{3}$ K,
but our results can easily be generalized to cases where this is
not so.

In the next section, we discuss the effects of \higgs variation
on BBN. In Sec.~{III}, we present the effect on
recombination and calculate the CMB temperature anisotropy.
Finally, we discuss our results and conclusions in Sec.~{IV}.
We take $\hbar = c = 1$ throughout.

\section{Effects on Big Bang Nucleosynthesis}

A time variation in \higgs affects Big Bang nucleosynthesis in several ways.
The variation of \higgs changes $G_F$, as given in equation (\ref{G_F}),
which alters the weak $n \leftrightarrow p$ interaction rates.
Changing \higgs also alters the fermion masses
as
\begin{equation}
m_F \propto \left<\phi\right>.
\end{equation}
Consequently, the electron and the quark masses change.
The change in the electron mass alters the $n \leftrightarrow p$ rates
(see below) and affects the evolution of the electron-positron energy
density during the epoch of nucleosynthesis.
Due to the
changes in the $u$ and $d$ quark masses, the neutron-proton
mass difference changes, and the pion mass varies.  The latter
affects nucleosynthesis through the change in the binding energy
of the deuteron.
We have incorporated all of these effects into our calculations.

\begin{figure}[tb]
\centerline{\epsfxsize=3.5truein\epsffile{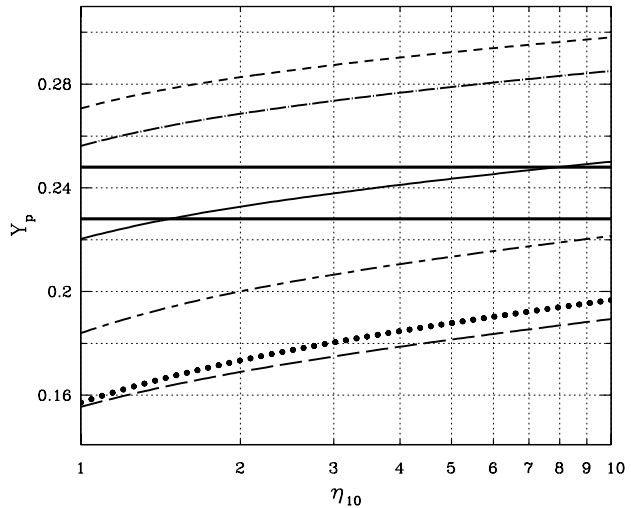}}
\caption{Effects on the $^4$He abundance ($\rm Y_p$) of a +5\% variation in \higgs.
The solid curve represents $\rm Y_p$ as a function of the baryon-photon ratio $\eta$
($\eta_{10}=10^{-10}\eta$) without \higgs variation.
Other curves show the isolated effects of changing $G_F$ (dot-dash)
the neutron-proton mass difference (long dash), the electron
mass (short dash) and the binding energy of the deuteron [with $r=6$ in equation
(\ref{Bdfit})]
(long dash - short dash)
The dotted curve gives $\rm Y_p$
when all the effects are considered together.  For reference,
the observational constraints on $\rm Y_p$ are shown as two horizontal
lines \cite{bbn}.}
\label{fig1}
\end{figure}

Consider first the neutron-proton mass difference,
$Q \equiv m_n - m_p$.
Following Ichikawa and Kawasaki \cite{ichikawa}, we take
\begin{equation}
Q=-0.76{\rm \ MeV}+2.053{\rm \ MeV}
{{\left<\phi\right>}\over{\left<\phi\right>_0}},
\end{equation}
where $\left< \phi\right>_0$ is the present Higgs vacuum expectation value.
This change in $Q$ alters the
ratio of neutron to proton abundances in thermal equilibrium:
\begin{equation}
{{n_n}\over{n_p}}=e^{-Q/kT}.
\end{equation}
The effect on the $^4$He abundance due to the change in $Q$ (alone) is illustrated
in Fig.~\ref{fig1}.  An increase in \higgs leads to an increase in
$Q$, giving a smaller equilibrium neutron-proton ratio, which produces
a smaller $^4$He abundance.

The weak interactions which interconvert neutrons and protons
are affected by both the change in $G_F$ and $m_e$.
These interactions are
\begin{eqnarray}
n+\nu_e &\leftrightarrow& p+e^-,\nonumber \\
n+e^+&\leftrightarrow& p+\bar \nu_e,\nonumber \\
n&\leftrightarrow& p+e^-+\bar\nu_e.
\end{eqnarray}
The total $n \rightarrow p$ rate is
\begin{eqnarray}
\label{n->p}
&&\lambda\left(n\rightarrow p\right)
={{4}\over {\pi^3}} n_nG_F^2 \int_{m_e}^\infty dE_e
{{E_e|p_e|}\over{1+{\rm \exp}[E_e/kT]}} \nonumber \\
&&\times \left\{ {{(E_e+Q)^2}\over{1+{\rm \exp}[-(E_e+Q)/kT_\nu]}}+
{{(E_e-Q)^2{\rm \exp}(E_e/kT)}\over{1+{\rm \exp}[(E_e-Q)/kT_\nu]}}
\right\} \nonumber \\
\label{rate}
\end{eqnarray}
where 
the subscripts $e$ and $\nu$ denote the quantities associated with the electron
and neutrino, respectively.
A similar expression can be derived for the $p \rightarrow n$ rate.
(Note that in the actual BBN calculation, the weak rates are scaled off of
the neutron lifetime; we have simply rescaled these rates using equation
\ref{n->p}).

Note that these rates depend on both $G_F$ and $m_e$, both of which
are altered by changing \higgs.  There is a further small change
in BBN produced by the effect on the expansion
rate of the total $e^+e^-$ energy density,
which depends on $m_e$. 
We have included this effect, although
it is small.  In Fig. 1, we show the effect on the
$^4$He abundance of changing $m_e$ and
$G_F$ separately.  Note that most of the change from $m_e$ is due
to the change in the weak rates, as we have noted, rather than
from the change in the expansion rate.  An increase in \higgs
results in a decrease in $G_F$, leading to earlier freeze-out
of the $n \leftrightarrow p$ reactions, producing more $^4$He.
Similarly, an increase in \higgs results in an increase in $m_e$,
decreasing the $n \leftrightarrow p$ reaction rates and also producing
more $^4$He.



Finally, changing \higgs alters the deuteron binding energy through
the change in the pion mass \cite{dixit_sher},
\begin{equation}
m_\pi^2 \propto m_u + m_d \propto \left<\phi\right>.
\end{equation}
Early calculations suggested a linear dependence of $B_D$ (the deuteron
binding energy) on $m_\pi$, and in previous discussions of BBN with a time-varying
\higgs, it was argued that this effect would be negligible and could be
ignored \cite{dixit_sher,ss} (although the effect of changing $B_D$ was explored
in a slightly different context in Ref. \cite{fairbairn}).

Although more recent calculations \cite{Bd1,Bd2} have not produced a definitive
result for the dependence of $B_D$ on $m_\pi$, they are in qualitative
agreement over our limited range of interest (i.e., very small
changes in $m_\pi$.)  In this regime, $B_D$ is a {\it decreasing}
function $m_\pi$ (the opposite of what was previously assumed
\cite{dixit_sher}).
The calculations in both \cite{Bd1} and \cite{Bd2} have very large
uncertainties, but within our narrow range of interest, we can safely
approximate the change in $B_D$ as a linear dependence on $m_\pi$,
through
\begin{equation}
\label{Bdfit}
{B_D \over B_{D_0}} = (r + 1) - r{m_{\pi} \over m_{\pi_0}}
\end{equation}
where the 0 subscript denotes the
values of these quantities at present, and the coefficient $r$ in
the fitting formula is to be derived
from Refs. \cite{Bd1,Bd2}.  We estimate the central values for $r$
from these references to be $r \approx 6$ \cite{Bd1} and
$r \approx 10$ \cite{Bd2}.  Although these values for $r$ are
quite different, we will see that they lead to similar constraints
on a change in \higgs.  On the other hand, our constraint
differs sharply from what is derived by ignoring the
change in $B_D$.

\begin{figure}[tb]
\centerline{\epsfxsize=3.5truein\epsffile{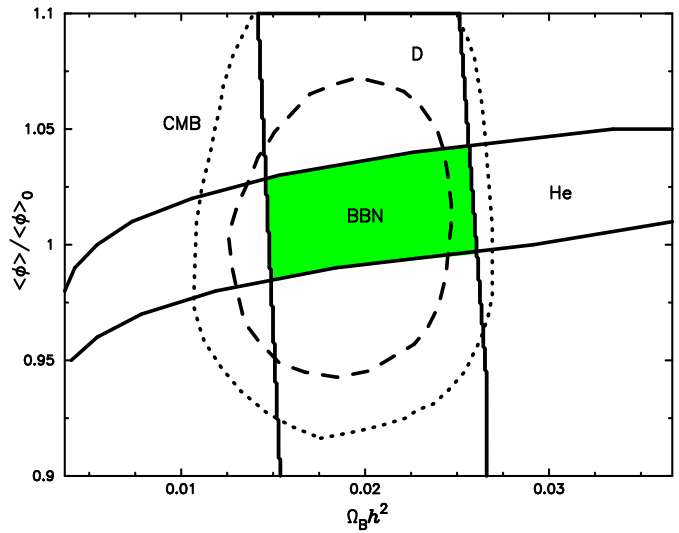}}
\caption{Regions consistent with the observational data.
The solid curves represent the constraints from the $^4$He and
deuterium abundances, assuming no change in the deuteron binding energy.
The region allowed by BBN is shaded.
The dotted and the dashed curves represent the 95\% and the 68\%
confidence level regions of the CMB observations, respectively.
}
\label{fig2}
\end{figure}

\begin{figure}[tb]
\centerline{\epsfxsize=3.5truein\epsffile{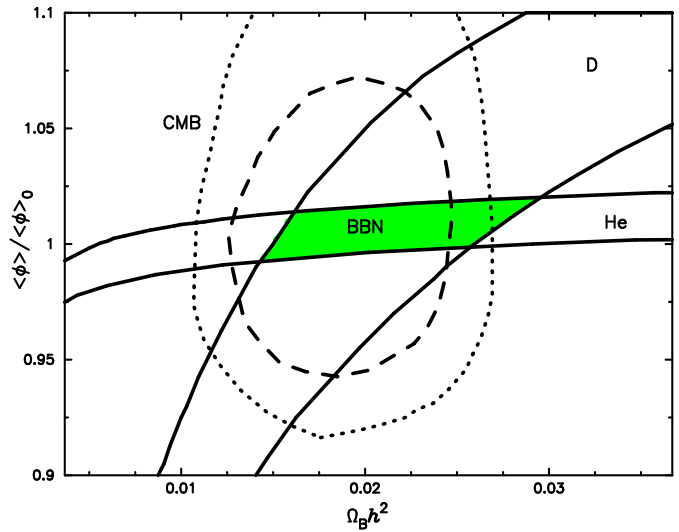}}
\caption{As Fig. 2, but the BBN limits now assume a change
in the deuteron binding energy given by equation (\ref{Bdfit})
with $r=6$.  The CMB contours are identical to those in Fig. 2.}
\label{fig3}
\end{figure}

\begin{figure}[tb]
\centerline{\epsfxsize=3.5truein\epsffile{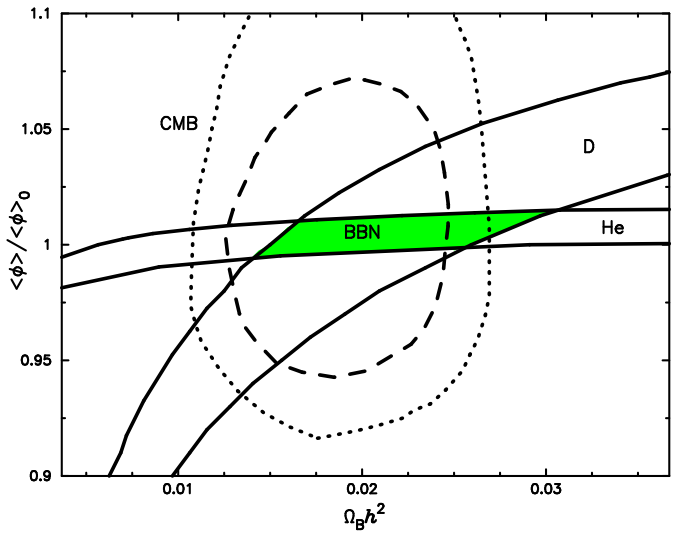}}
\caption{As Fig. 3, with $r=10$.  The CMB contours are identical to those in Figs. 2
and 3.}
\label{fig4}
\end{figure}

The effect on the helium abundance of changing $B_D$ is
shown in Fig. 1 (for $r = 6$).  It is clear that this
effect is not negligible.  Further, (and unlike
the other effects we have considered) changing $B_D$ has
a significant effect on the primordial deuterium abundance.
This is most evident from our graphs of the allowed region
for \higgs (see Figs. $\ref{fig2} - \ref{fig4}$ below).
An increase in \higgs leads to an increase in $m_\pi$, resulting
in a decrease in $B_D$.  This leads to a smaller equilibrium deuterium abundance.  Thus, the production of $^4$He begins
later, leading to a smaller $^4$He abundance, as shown in Fig. 1.
Somewhat paradoxically, a decrease in the deuterium binding energy
leads to an {\it increase} in the final deuterium abundance;
nucleosynthesis at lower temperatures faces a larger Coulomb
barrier, leaving more deuterium behind.

To derive BBN limits on a change in \higgs, we assume two free
parameters, the baryon
to photon ratio $\eta$,
and
$\left<\phi\right>/\left<\phi\right>_0$.
We vary these parameters within the range $\eta = 10^{-10} - 10^{-9}$
and 
$\left<\phi\right>/\left<\phi\right>_0 = 0.9 - 1.1$.

To derive our constraints, we consider only the abundances of
deuterium and $^4$He.
Recent observations \cite{bbn} suggest the limits
\begin{equation}
(\rm D/H)=3.0^{+1.0}_{-0.5}\times10^{-5},
\end{equation}
on the ratio of deuterium to hydrogen, and
\begin{equation}
\rm Y_p=0.238\pm0.005,
\end{equation}
on the primordial mass fraction of $^4$He.
We ignore the somewhat more uncertain limits on $^7$Li
in determining our constraints, although we have
verified that
the allowed regions we derive from BBN also
produce an acceptable $^7$Li abundance.

Our limits on shown in Figs. $\ref{fig2}-\ref{fig4}$,
where we have translated our constraints on $\eta$ into
limits on $\Omega_B h^2$.
In Fig. \ref{fig2} we show (for reference) the
effect of neglecting the deuteron binding energy,
while Figs. \ref{fig3} and \ref{fig4} correspond to
our two choices for $r$.  The solid lines give the constraints
from the deuterium and $^4$He abundances, with the shaded area
indicating the region producing an acceptable abundance of
both.

The importance of including the effect of the deuteron binding
energy is obvious from comparing Fig. \ref{fig2} to
Figs. $\ref{fig3} - \ref{fig4}$.  In Fig. \ref{fig2}, the
regions of allowed deuterium and allowed $^4$He are almost
parallel to the axes, indicating that the deuterium abundance
is nearly independent of changes in \higgs; in this approximation,
the abundance of deuterium sets the bounds on $\Omega_B h^2$, with
the $^4$He abundance then determining the limits on \higgs.
In Figs. $\ref{fig3} - \ref{fig4}$, we see a strong dependence
of the deuterium abundance on \higgs (through the deuteron
binding energy), while the allowed band for $^4$He is much narrower.
On the other hand, Fig. \ref{fig3} and Fig. \ref{fig4} are 
reasonably similar, allowing us to put useful constraints on the time
variation in \higgs despite the uncertainty in the dependence
of $B_D$ on $\left<\phi\right>$.

Including the change in the deuteron binding energy,
we find that the BBN constraint on \higgs variation is extremely
tight.  Defining $\Delta \left<\phi\right> \equiv \left<\phi\right> - \left<\phi\right>_0$, we get
\begin{equation}
- 0.7 \times 10^{-2} < \Delta\!\!\left<\phi\right>\!/\!\left<\phi\right>_0
< 2.0 \times 10^{-2},
\end{equation}
for $r = 6$, and
\begin{equation}
- 0.6 \times 10^{-2} < \Delta\!\!\left<\phi\right>\!/\!\left<\phi\right>_0
< 1.5 \times 10^{-2},
\end{equation}
for $r = 10$.

\section{Effects on the cosmic microwave background radiation}
Since the weak interaction has no effect on
the recombination process,
the only effect of changing \higgs arises through the change
in the electron mass, $m_e$ \cite{kujat}.  (We
have neglected any change in the mass of the dark matter particle.
For axions, we expect no change, while the change in the
mass of a supersymmetric dark matter particle will be model-dependent).
The change
in $m_e$ alters both the Thomson scattering cross section and
the binding energy of hydrogen.
The Thomson scattering cross-section
is
\begin{equation}
\sigma_T={{8\pi\alpha^2}\over{3}}
m_e^{-2},
\label{thomson}
\end{equation}
while
the binding energy of hydrogen is given by
\begin{equation}
B={{\alpha^2}\over{2}}
m_e.
\label{hydrogen}
\end{equation}

The ionization fraction, $x_e$, is determined by the balance
between photoionization
and recombination.
The evolution of $x_e$ with a variation in \higgs
is given by \cite{kujat,recombination,ionization}
\begin{equation}
-{{dx_e}\over{dt}}={\sc C'}\left[ {\sc R'}n_p x_e^2-\beta'
(1-x_e){\rm exp}\left(-{{B'_1-B'_2}\over{kT}}\right)\right],
\end{equation}
where $\sc R$, $\beta$, $B_n$, and $\sc C$
represent the recombination coefficient, the ionization coefficient,
the binding energy of the $n_{th}$ hydrogen atomic level, 
and the Peebles correction factor, respectively.
The primed quantities stand for the modified coefficients following \higgs
variation \cite{kujat}.
Finally, the differential optical depth of the CMB photons is determined by
\begin{equation}
\dot\tau= x_e n_p \sigma_T.
\end{equation}

\begin{figure}[tb]
\centerline{\epsfxsize=3.5truein\epsffile{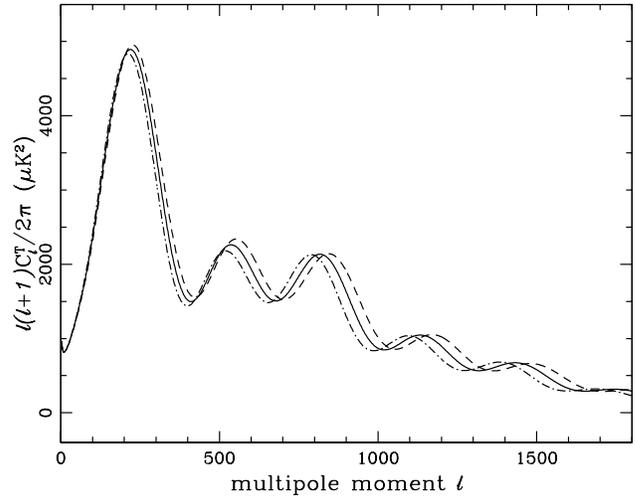}}
\caption{The effect of \higgs variation on the
CMB temperature anisotropy power spectrum for the $\Lambda$CDM
concordance model of Ref. \cite{wang} ($\Omega_\Lambda = 0.66$,
$\Omega_{CDM}h^2 = 0.12$, $\Omega_B h^2 = 0.2$, $n_s = 0.91$).
The solid line represents the power spectrum
without \higgs variation. The dashed and the dot-dashed lines correspond to
5\% and $-5$\% variations of \higgs, respectively.}
\label{fig5}
\end{figure}

With these modifications, we use CMBFAST \cite{cmbfast} to generate the
theoretical power spectra of temperature fluctuations.
(See Ref. \cite{kujat} for a more detailed discussion).
Fig.~\ref{fig5} shows the effects of \higgs variation
on the CMB power spectrum.  An increase in \higgs leads to a
decrease in the Thomson scattering cross section and an increase
in the hydrogen binding energy.  The latter effect is the most important,
since it shifts the last-scattering surface to a smaller comoving
length scale, moving the curve in Fig. \ref{fig5} to the right.

Using RADPACK \cite{radpack},
we perform a $\chi^2$ analysis for the most recent CMB 
data: VSA, BOOMERANG, DASI, MAXIMA and 
CBI \cite{vsa,boom,maxima,dasi,cbi}.
Assuming a flat universe 
($\Omega_{\rm B}+\Omega_{\rm CDM}+\Omega_\Lambda=1$),
we set the model parameter space 
$0.5\le h_0 \le 0.8$ and $0.3\le\Omega_M\le0.4$
(where $\Omega_M \equiv \Omega_B + \Omega_{CDM}$),
and we take the spectral index $n_s$ to lie in the range
$0.7 \le n_s \le 1.3$.
The power spectra are obtained
by a modified CMBFAST with \higgs variation.
We then marginalize over \higgs and $\Omega_{\rm B}h^2$.

The CMB constraints are shown in Figs. $\ref{fig2} -
\ref {fig4}$ (identical in all three figures). The dotted and
the dashed lines represent the 95\% and the 68\% confidence level regions
of the CMB experiments, respectively.  In general, the CMB calculations
provide less stringent constraints than BBN; almost the entire region
allowed by BBN lies inside the 95\% confidence region from the CMB.
The small area allowed by BBN and excluded by the CMB represents
a tighter bound on $\Omega_B h^2$, but it does not alter the limits
on \higgs.

\section{discussion}

This work provides new constraints on the possible time-variation
of \higgs from both BBN and the CMB.  We find that the BBN
limits are significantly more stringent than those that can
be derived from the CMB.  More specifically, we get
\begin{equation}
-(0.6 - 0.7) \times 10^{-2} < \Delta\!\!\left<\phi\right>\!/\!\left<\phi\right> <
(1.5 - 2.0) \times 10^{-2},
\end{equation}
with the exact numbers depending on the assumptions made about the dependence
of the
deuteron binding energy on $m_\pi$.  These limits are considerably more
stringent than those in, e.g., Ref. \cite{ss}, due to better observational
limits on the primordial element abundances, as well as the inclusion
of the (previously-neglected) effect of \higgs on the deuteron
binding energy.

In contrast, the limits from the CMB are much less stringent.  This
conclusion is similar to the results obtained for the time-variation
in $\alpha$ obtained by Avelino et al. \cite{avelino}.
(Note also that the effect of a time-variation in $\alpha$ and a time variation
in \higgs are almost completely degenerate \cite{kujat}, but
BBN breaks this degeneracy).
The chief advantages
of considering the CMB limits are that the calculation is much
more straightforward (dependent only on the change in the electron mass),
and that the CMB limits should improve sharply in the near future as
more observational data comes in.
Of course, if \higgs varied between the epoch of BBN ($T \sim 10^{10} - 10^9$
K) and the epoch of recombination ($T \sim 10^3$ K), then the limits
we have derived could be applied separately to constrain
$\Delta\!\!\left<\phi\right>\!/\!\left<\phi\right>$ at each epoch.

\section*{Acknowledgments}
We are grateful to G. Steigman, R. Furnstahl, M. Tegmark,
J. Kneller, and S. Raby for helpful discussions.
We thank U. Seljak and M. Zaldariagga for
the use of CMBFAST \cite{cmbfast} and L. Knox for the use
of RADPACK \cite{radpack}.  J.J.Y. was supported
by a University Fellowship from The Ohio State University.
R.J.S. was supported by the Department
of Energy (DE-FG02-91ER40690).

\end{document}